# Businesses in high-income zip codes often saw sharper visit reductions during the COVID-19 pandemic

Aditya Kulkarni, Department of Computer Science, Yale University, New Haven, USA

Min Kim, Department of Economics, Trinity University, San Antonio, USA

Joydeep Bhattacharya, Department of Economics, Iowa State University, Ames, USA

Jayanta Bhattacharya, Department of Medicine, Stanford University, Stanford, USA

**Abstract**

As the COVID-19 pandemic unfolded, the mobility patterns of people worldwide changed drastically. While travel time, costs, and trip convenience have always influenced mobility, the risk of infection and policy actions such as lockdowns and stay-at-home orders emerged as new factors to consider in the location-visitation calculus. We use SafeGraph mobility data from Minnesota, USA, to demonstrate that businesses (especially those requiring extended indoor visits) located in affluent zip codes witnessed sharper reductions in visits (relative to pre-pandemic times) outside of the lockdown periods than their poorer counterparts. To the extent visits translate into sales, we contend that post-pandemic recovery efforts should prioritize relief funding, keeping the losses relating to diminished visits in mind.

**Competing interests:** The authors declare no competing interests.

**Author Contribution:** The paper was jointly conceived by all four authors. The data analysis was done mainly by Kim and Kulkarni, and all four authors did the writing.

**Data availability statement:** The data supporting this study's findings are available from SafeGraph (www.safegraph.com) upon academic request.

**Code availability statement**: No custom code was used to generate the results. Any code used will be made available upon e-mail request to any of the authors.

## Introduction

The urban landscape, a vital engine of social and economic opportunity, is a daily witness to intense face-to-face socioeconomic interactions that exhibit excellent regularity[a]. In the absence of restrictions on spatial mobility, residents of a city typically have plans for times within the day that they wish to stay indoors vs. outdoors or which businesses or locations to visit. However, the same people are forced to re-optimize once constraints are externally imposed due to a disease-related lockdown.[b] Depending on the timing and nature of the restrictions, some residents a) cut down most travel outdoors and hunker down at home, b) reduce mainly non-essential or easily substitutable travel, and c) switch to other businesses and locations different from ones they would patronize pre-lockdown.[c,d] Businesses, especially those relying on face-to-face transactions, see the effects of changes in the mobility calculus on the number of visits they see.

In March 2020, the transmission of the Sars-Cov2 virus had reached pandemic proportions, and governments worldwide were scrambling to formulate a non-pharmaceutical policy response. Wherever possible, governments enacted aggressive policies, including "shelter in place" and emergency closures of all non-essential services, with associated economic and social consequences. The first stay-at-home order was imposed in Minnesota, USA, on March 16, 2020[1]. It forced many Minnesotans to remain home most of the time and re-optimize their daily outdoor-indoor mobility calculus. Some had to cut down on restaurant visits, others on trips to movie theaters, doctor offices, pharmacies, or grocery stores. How did the vastly altered mobility landscape affect businesses and point-of-interest locations?[e] Did location matter: did companies located in high-income zip codes see sharper declines in visits (relative to pre-pandemic levels)? Did timing matter, whether it was a lockdown period or not? This paper documents pandemic-influenced visiting patterns in businesses in Minnesota's high- and low-income zip codes following the onset of COVID-19.

The research is framed as follows. Consider visits by potential customers to two otherwise-identical businesses selling the same product, one located in a high and the other in a low-income neighborhood. Suppose, pre-pandemic, the business located in the high-income area saw more visits than its counterpart in the low-income area. We want to know, how does the number of visits change post-pandemic? If visits fell, did they fall more (and stayed that way longer) in the high-income neighborhood store? How about visit patterns to businesses selling other products? Other neighborhoods?

Finally, why are we interested in these questions? We wish to know how non-pharmaceutical interventions, designed to contain the spread of the disease, affected businesses based on their location in high or low-income areas. This focus on the location of the business in a high or low-income locality is novel and allows the research to offer guidance on post-pandemic governmental relief allocation.

Ideally, we would like to know how the changed visitation calculus generally affected sales or GDP. Before proceeding further, it bears emphasis that we do not have access to expenditure data, i.e., we do not know the identity of the visitors nor how much was spent or earned due to a single visit to any location. For us, visits are an intermediate marker of economic activity. While visits to a store in normal times closely track sales, that correlation may have weakened during the pandemic, especially if sales moved online, businesses started delivery programs, or an UberEats driver executed multiple restaurant pickup orders in a single visit. [f]

To place our research questions in context, consider Figure 1, which compares the mobility surrounding restaurants within two cities in Minnesota with vastly different median incomes: Prior Lake has a median household income of $109,609, more than double Hibbing's income of $49,009. The vertical axis measures restaurant visits during 2020-21 as a fraction of the same for 2019. (The data sources are described further below.)

As 2020 starts, it is apparent that both the yellow (Prior Lake) and the blue (Hibbing) lines are similar through January and February, with the yellow above the blue. However, soon after the first lockdown period started in March 2020, both lines began to show steep declines. (Note that the shaded regions represent indoor dining venues' lockdown periods.) Once restaurants reopened in June 2020, differences between the two came into sharp focus: Hibbing's restaurant visits recovered to near pre-pandemic levels, while those in Prior Lake restaurants only achieved a sluggish recovery and remained at depressed levels for months after. Prior Lake's restaurants show a significant spike in visits during the second lockdown, possibly due to more curbside pickup visits for Thanksgiving. In contrast, Hibbing's restaurants slowly lose visitors. Hibbing's restaurants likely relied more on regular in-person visit traffic since a greater proportion of their visits were of long duration before the second lockdown.

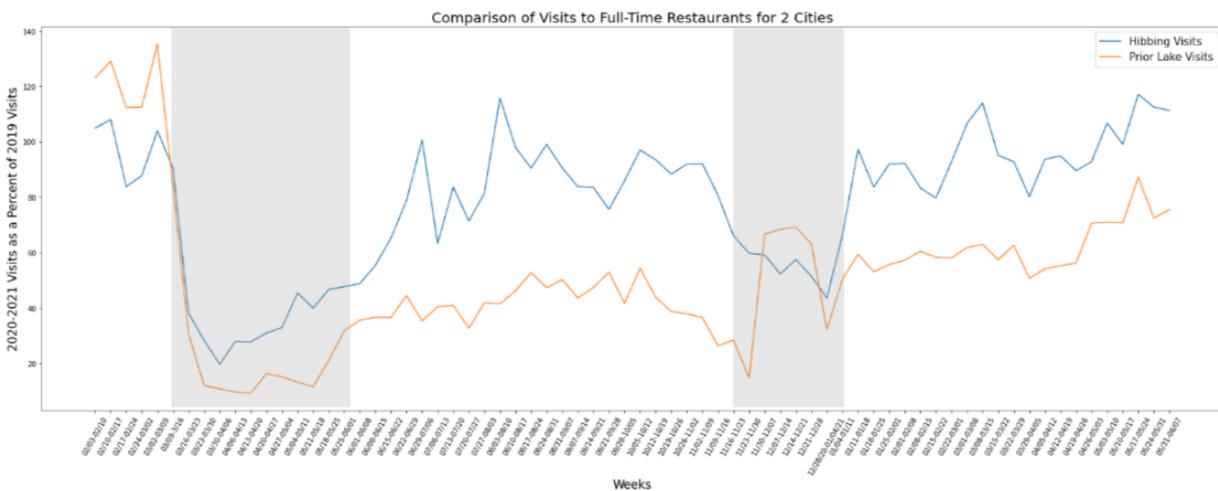

*Figure 1 Comparison of visits to full-time restaurants for Hibbing and Prior Lake Cities*

Our primary contributions are as follows:

- We outline a pairwise comparison method that compares the reduction in visits to businesses in two groups of zip codes (top and bottom one-third of zip codes by median income). This method accounts for differences in individual zip code population sizes and calibrates it to the usual number of visits to each business location based on past years.

- The main table shows twenty-one different North American Industry Classification System (NAICS) business categories where businesses in wealthier zip codes had a more significant decrease in visits than businesses in poorer zip codes. These metrics are calculated for periods in different stages of the pandemic. The blue cells (indicating a higher reduction in the poorer zip codes) are concentrated in the lockdown period. In comparison, the red (higher reduction in the rich zip codes) cells are concentrated in businesses that encourage extended indoor visits (sit-down restaurants, religious organizations, and movie theaters) throughout the pandemic (not just during the lockdown). In low-income zip codes, the drop in the number of visits to all kinds of businesses during the lockdown was more pronounced than in high-income zip codes.

- This paper contains kernel density estimation plots of all possible pairwise comparisons of changes in visits for select business categories to show the dispersion of comparisons during different periods.

Our research is not intended to shed light on the measures required to contain the spread of the disease[2,3]; it takes the disease, its progression, and the policies as given and attempts to understand their effect on commerce. The answers we find could be of some value to policymakers trying to design non-pharmaceutical responses to epidemics under twin objectives, a) contain or reduce disease spread, *and* b) generate minimal disruption to people's lives and business operations. If the aim is singularly to contain the epidemic, then indiscriminate lockdowns are probably the best policy response. Our work suggests that if objective (b) is taken seriously, the knowledge that businesses in rich and poor urban areas withstand mobility shocks differently can help decide which companies and locations to place under lockdown and which to spare.

In a broader context, our work documents one aspect of the significant changes in urban interactions due to social distancing measures, even as cities embrace new strategies to address inequities caused by such measures.[g] Limited by a lack of spending/revenue data, our research offers tentative guidance on where scarce pandemic-relief dollars should go. For example, suppose movie theaters are visited mainly by the affluent, who sharply curtail theater visits (but not pharmacy visits) post-lockdown. In that case, it could be argued relief dollars should go to theater owners before going to pharmacy owners. Bear in mind businesses witnessing the sharpest reductions in visits are often places that hire many low-wage workers.

Our paper aligns with recent literature analyzing mobility differences in the U.S., finding that high-income neighborhoods increased days at home substantially more than low-income neighborhoods[4–6]. Some of this is explained by the enhanced capacity of the rich to work from home. Another angle that may explain the difference in stay-at-home days is risk perception[7,8]. Bonaccorsi et al. (2020)[9] find that Italian municipalities richer in terms of social indicators and fiscal capacity are more affected by the loss in mobility efficiency in the aftermath of the lockdown. Indeed, they report two seemingly opposite patterns: Individual indicators (average income) show that the poorest are more exposed to the economic consequences of the lockdown; conversely, aggregate indicators (municipality level), deprivation, and fiscal capacity, reveal that wealthier municipalities are those more severely hit by mobility contraction induced by the lockdown. They do not, however, study the effect on businesses.

## Data

### i.    Mobility Pattern

We use data from the SafeGraph COVID-19 Data Consortium to observe human mobility patterns. SafeGraph collects GPS information from about forty-five million anonymized smartphone devices (10% of devices) and 3.6 million point-of-interest (POI) locations in the U.S[h,i]. The GPS data comes from apps where users have consented to location tracking. Within the SafeGraph COVID-19 Data Consortium, there are three primary datasets, of which we use two: Weekly Patterns and Core Places. From the Weekly Patterns dataset, we use data on the number of visits each week to each POI in the area we are analyzing. SafeGraph counts a visit to the POI by checking if the GPS location matches the inner boundary of a POI location[j]. We use the Core Places dataset to bring additional information on each POI, such as the North American Industry Classification System (NAICS) code[10], street address, city, region, and zip code. We merge both datasets by matching the unique POI identifiers "safegraph_id" (to separate POIs of the same name but different locations) to create a single dataset with millions of records. Each record contains the POI name, the two dates marking the beginning and end of a specific week, the number of visits in that week, the POI's city, the POI's zip code, and the POI's NAICS code, and the POI's "safegraph_id."

### ii. Income and population

Income and population data comes from the 2015-2019 American Community Survey 5-year Estimates[11]. The zip code is our geographic unit in this study. To divide rich and poor areas, we rank zip codes by the median income of residents; the top one-third are classified as high-income, and the bottom one-third as low-income.

## Methodology

We chose to analyze the difference in mobility patterns in Minnesota's businesses because of the range of clearly articulated policies employed during various pandemic stages.

The first stage is the pre-pandemic baseline, from February 2, 2020, to March 16, 2020; the second is the first lockdown phase, March 17, 2020, to June 1, 2020; the third is the interim when businesses reopened, June 2, 2020, to November 23, 2020. The next phase is the second lockdown, November 24, 2020, to January 11, 2021, and the last is the reopening period, from January 11, 2021, to May 31, 2021. These dates line up with the initial plans for reopening Minnesota.

Our geographical unit is the zip code, denoted by *j*. Let *i* represent the business category (e.g., full-time restaurants), *k*, the number of businesses under category *i*, and *t* the period mentioned above. Let *v* denote visits, our measure of mobility. First, we compute the number of visits to a given business in a zip code, $v_{ijkt}$. Next, we compute $V_{ijt} \equiv \sum_k v_{ijkt}$, the sum of visits to every business in category *i* in zip code *j*. Then, we compare $V_{ijt}$ against the number of visits in the corresponding weeks of the year 2019, denoted by $V_{ijt_{pre}}$ [k]. Comparing the number of visits in 2019, we can account for normal variations such as holidays or seasonality. In addition, it removes the tracking bias consistent across 2019-2021 in the same area.

Table 1 shows the corresponding period dates from 2020-2021 to 2019.

|  | **Year 2019** | **Year 2020-2021** |
|---|---|---|
| Normal | 2/04/19 - 3/18/19 | 2/03/20 - 3/16/20 |
| Lockdown 1 | 3/18/19 - 6/03/19 | 3/16/20 - 6/01/20 |
| Reopening 1 | 6/03/19 - 11/25/19 | 6/01/20 - 11/23/20 |
| Lockdown 2 | 11/25/19 - 20 - 1/14/19 | 1/23/20 - 1/11/21 |
| Reopening 2 | 1/13/19 - 6/01/19 | 1/11/21 - 5/31/21 |

*Table 1 Corresponding Period Dates*

Our identifying assumption is, had the pandemic not happened, $V_{ijt}$ would be very similar to the same time period in 2019, $V_{ijt_{pre}}$. This assumption allows us to ascribe changes in visits to the lockdown, the fear of the virus, or both. We divide the number of visits by the zip code population to make changes comparable across zip codes.

In our sample, $j = \{1,2,..,574\}, j \epsilon\ H \cup L$ where 287 zip codes are classified as low-income (denoted $L$), and the rest are high-income (denoted $H$). Next, we compare all possible (82,369 combinations) pairwise differences in the change in the number of visits for high- and low-income zip codes. That is, we take one high-income zip code and compare its population-deflated $\Delta_{ij} = \left(V_{ijt} - V_{ijt_{pre}}\right)_{j \in H}$ with the same for a low-income zip code $\Delta_{ij} = \left(V_{ijt} - V_{ijt_{pre}}\right)_{j \in L}$. We do this for all possible pairwise combinations for all $j$.

$$\left(V_{iht} - V_{iht_{pre}}\right) - \left(V_{ilt} - V_{ilt_{pre}}\right) \text{ for } \forall h \in H \text{ and } \forall l \in L$$

If this number is negative, it follows that visits to business $i$ during $t$ in high-income zip codes saw a sharper reduction compared to pre-pandemic levels in low-income zip codes.

Finally, we construct the distribution of all differences across all possible combinations. We produced such distributions for $i = \{1,2,,...21\}$ different business categories and $t = \{1,..,5\}$ different periods. The results are summarized in Table 2.[1]

Consider the following simple illustration: imagine two rich areas, $a$ and $b$, and two poor areas, $x$ and $y$. For each business type, there are four possible pairwise comparisons. For business-type $i$, the possible comparisons are $\Delta_{ia} - \Delta_{ix}$, $\Delta_{ia} - \Delta_{iy}$, $\Delta_{ib} - \Delta_{ix}$, $\Delta_{ib} - \Delta_{iy}$. $\Delta_{ia} - \Delta_{ix} < 0$ means visits to business $i$ in area $a$ decreased more than in area $x$ compared to visits in 2019. Similarly $\Delta_{ia} - \Delta_{ix} = 0$ means the change in visits is identical in areas $a$ and $x$. We obtain the distribution by computing this pairwise comparison for all possible cases. A distribution centered at zero means that the change in visits in low- and high-income areas is identical.

## Results: Kernel Density Plots

Figure 2 and Figure 3 plot the kernel densities (KDE) derived from a smooth, continuous density estimation of all possible pairwise comparisons of changes in visits to full-time restaurants and groceries between low- and high-income zip codes. Each color represents the distribution during the indicated period.

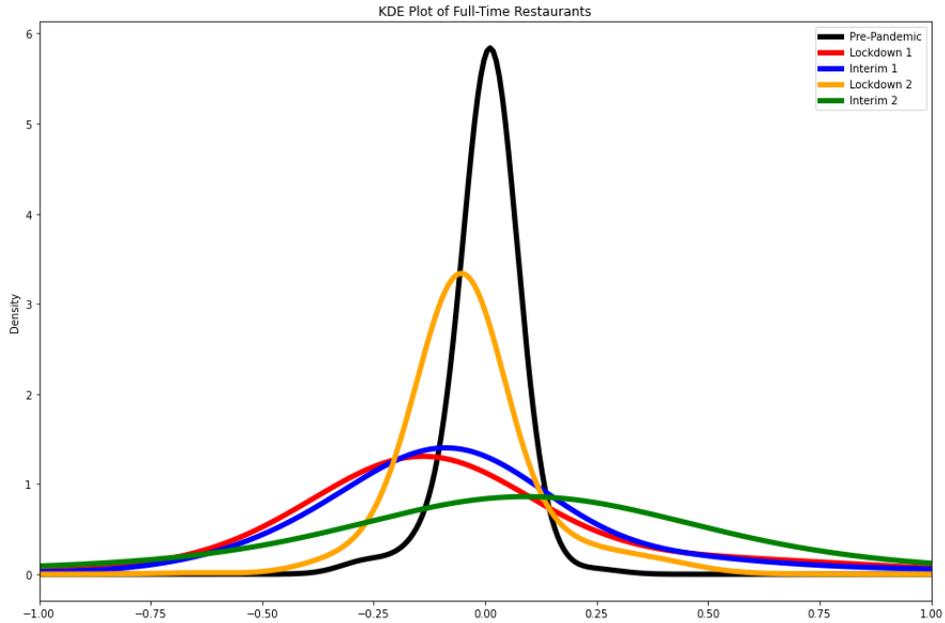

*Figure 2 Relative changes in the number of visits to full-time restaurants between low- and high-income zip codes*

The black plot is the pre-pandemic difference centered at zero for both business locations. The variance is slight, meaning visits to full-time restaurants or groceries stayed the same from the previous year by location. In other words, low- and high-income areas had a parallel trend in visits in the pre-pandemic period. The same trend is observable for almost all categories, $i$. Things changed once the pandemic got underway. The red line plotted the distribution during the first lockdown in March 2020: now, the distribution is centered to the left of zero, meaning, on average, restaurants in prosperous zip codes had fewer visitors than businesses in low-income zip codes. This pattern persists until the last reopening period, starting in 2021.[m]

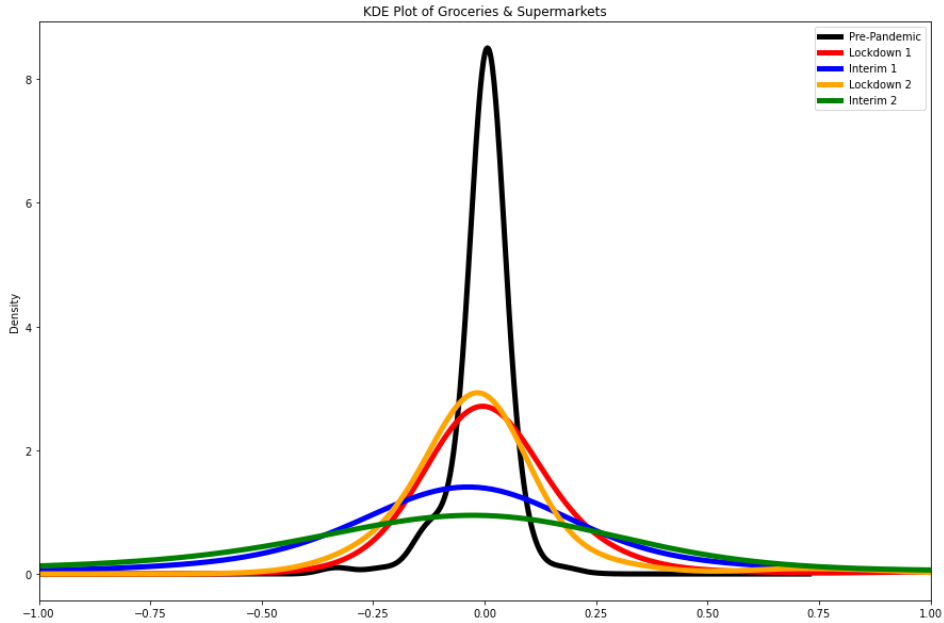

*Figure 3 Relative changes in the number of visits to groceries and supermarkets between low- and high-income zip codes*

On the other hand, the pattern of visits to groceries and supermarkets in Figure 3 is similar in both zip codes across all periods. The distribution is centered on zero, indicating both the rich and the poor zip codes reduced visits in much the same way. One way to explain this finding is to classify visits to groceries and supermarkets as essential-oriented and relatively unaffected by income differences across zip codes. People in both zip codes demand fresh produce, milk, and other perishables. The time gap between visits may have changed after the pandemic started, but the number remained the same. Under this classification, visits to full-time restaurants are not essential-oriented.

## Results: pairwise comparisons frequencies, Table 2

Table 2 shows how the pairwise comparisons are distributed using $(V_{iht} - V_{iht_{pre}}) - (V_{ilt} - V_{ilt_{pre}})$. For example, consider a city with three high-income zip codes (A, B, and C) and three low-income zip codes (1,2, and 3). Not knowing whether A is a representative zip code, we take every zip code and construct nine pairwise comparisons: (A, 1), (A, 2), (A, 3), (B, 1), (B, 2), (B, 3), (C, 1), (C, 2), (C, 3). Next, we compare the change in visits (population deflated) to locations in low-income zip codes to high-income ones. It stands to reason the number of visits will be different across zip codes. To make the comparison fair, we compare changes to their pre-pandemic levels. Therefore, each pairwise comparison represents a *double difference*; 1) difference in visits pre- and post-pandemic in each zip code, 2) difference across low- and high-income zip codes.

Each cell in Table 2 is a number capturing the *share* of the entire set of pairwise comparisons that show negative changes. For example, 66.6% means the change is negative in six out of the entire set (nine) of pairwise comparisons. Specifically, in six out of nine pairwise matches, visits to locations in high-income zip codes fell more (relative to pre-pandemic levels) than in low-income zip codes. The entire distribution is centered on a negative number in this specific example. Had the decline in visits been similar across low- and high-income zip codes, the distribution (for large samples) would be centered on zero: roughly 50% of pairwise matches would show a negative number, and the remaining 50% would show a positive number. A number *greater* than 50% indicates that the distribution is centered to the left of zero. This means, in this instance, compared to pre-pandemic times, locations in high-income zip codes showed a more significant drop in visitors than did low-income zip codes. We colored cells dark red if the percentage exceeds 70%, light red for greater than 60% and less than 70%, light blue for greater than 30%, less than 40%, and dark blue for less than 30%. Thus, blue indicates that poorer zip codes saw a sharper reduction in visits to that type of service than rich zip codes, and the color red, vice versa.

Table 2 shows that people in more prosperous zip codes reduced visits to restaurants, religious organizations, and movie theaters. On the other hand, poorer zip code people reduced visits to plumbing, heating & A.C. contractors, veterinary services, funeral homes, police stations, and libraries. One notable feature of Table 2 is that many blue cells are for the lockdown period, implying poorer zip code-people people substantially reduced their mobility during the first lockdown. Another noticeable feature is that more affluent people stopped visiting places known to be prone to infection, suggesting better abilities to comply with the lockdown policy via switching to substitutes. For example, fewer restaurant visits could be because more prosperous people started using more delivery services. The change in visits to essential services such as supermarkets, gasoline stations, and medical services is similar for rich and poor areas. The number of rich and poor zip codes with at least one location of a particular business category are listed in the last two columns within the Appendix, Table A1. These two columns shed light on the distribution of a particular business type between rich and poor zip codes (e.g., there is at least one childcare business location in 149 of the rich zip codes compared to only 73 of the poor zip codes).

A notable fact from Figures 1 and 2 is that dispersion in mobility increased during the pandemic. Compared to the pre-pandemic level, the variance in the change in visits across rich and poor zip codes increased regardless of the type of business. This is observable even in industries, say groceries, where the distribution of the difference post-lockdown is centered at zero. This means that, on average, the reduction of visits to groceries is similar between the rich and poor areas. Still, people within similar income groups started to exhibit different mobility patterns after the pandemic. For example, some affluent areas sharply reduced outdoor trips compared to others from similar income areas. Some poor areas maintained a level equivalent to the pre-pandemic level, while others declined to go outside. This implies that factors other than median income may affect the rich and poor areas to behave differently. These could be income inequality, the number of jobs that can be done remotely, the age composition of the zip code, the perception of the epidemic, and so on.

To sum up, before the pandemic, visits to certain businesses were similar regardless of the median income of the zip code – the pre-pandemic distribution of the pairwise comparison is centered at zero. The slight variance of the pre-pandemic distribution indicates that the mobility patterns were homogeneous within the income group. However, rich and poor areas diverged into different visit patterns after the pandemic. This is confirmed because the distribution after the pandemic moved in different directions depending on the type of service. Furthermore, higher pairwise distribution variance during the pandemic period suggests that people diverge from the typical income group pattern after the pandemic. This disparity could be explained by factors other than income.

## Conclusion

As shown above, the COVID-19 pandemic and subsequent restrictions impacted different types of businesses in zip codes with high and low median incomes. We show that the effects differed across various business categories in distinct periods. Comparing the number of visits during the pandemic to data from 2019, more than a year before the COVID-19 pandemic hit the U.S., and adjusting visit totals relative to current population counts in each zip code were factors we used to normalize data into a comparable format. Regardless of how businesses were hit, there was a considerable divergence in mobility between richer and poorer zip codes in most business categories. In addition, within each income group, the mobility of people diverged from the pre-Covid pattern of the group.

We recognize that externally defined job classifications, such as the "essential worker" category, may have contributed to our observed visit patterns. This research can assist policymakers in implementing lockdown policies but still want to keep economic disruptions to a minimum. However, our analysis is based on changes in the number of visits, not spending. Although the number of visits is a good proxy for revenue, it still has limitations. Furthermore, because there was a different pattern of visit changes across zip codes of high and low median income, the results can contribute to the design of post-pandemic recovery programs.

A few caveats are important to record. First, when it comes to restaurant and grocery-store visits, it bears mention that residents of the more affluent neighborhoods are more likely to have used delivery options such as UberEats. Additionally, an UberEats driver could have executed multiple orders to a restaurant in one visit. Second, the lack of data on expenditures is a significant limitation. For example, residents in low-income neighborhoods may be unable to buy certain items in bulk (hence, have to make more visits) compared to their counterparts in more affluent areas.

Future research could explore how mobility changed based on various metrics, such as the most common jobs held by people in different zip codes. It could also look at changes in people's spending or modes of transportation, spending patterns, business revenue, and associated job losses to assess the economic impact of the policy. Lastly, there is also potential for different mobility patterns between richer and poorer zip codes in states other than Minnesota due to various states' unique approaches to mitigating COVID-19 exposure.

| Main Category | Name | 2/03/20 - 3/16/20 | 3/16/20 - 6/01/20 | 6/01/20 - 11/23/20 | 11/23/20 - 1/11/21 | 1/11/21 - 5/31/21 |
|---|---|---|---|---|---|---|
| | Percentage of pairwise comparison: $(V_{iht} - V_{iht_{pre}}) - (V_{ilt} - V_{ilt_{pre}}) < 0,$ for $\forall h \in H$ and $\forall l \in L$ | | | | | |
| Food Services | Full-Service Restaurants | 44.7% | 74.5% | 66.7% | 68.3% | 58.9% |
| | Limited-Service Restaurants | 56.1% | 55.7% | 60.5% | 49.3% | 55.1% |
| Essential Goods/Services | Meat, Seafood, & Fruit/Vegetable Markets | 47.6% | 34.1% | 50.8% | 50.7% | 49.2% |
| | Supermarkets & Grocery Stores | 42.3% | 48.9% | 60.4% | 58.5% | 57.8% |
| | Gasoline Stations | 52.7% | 57.5% | 56.1% | 59.0% | 43.7% |
| | Plumbing, Heating, & A.C. Contractors | 43.5% | 41.1% | 38.4% | 37.8% | 55.0% |
| Medical Locations | Offices of Physicians | 46.0% | 56.8% | 59.4% | 54.6% | 54.6% |
| | General Medical & Surgical Hospitals | 43.7% | 54.6% | 55.3% | 48.2% | 46.3% |
| | General, Surgical, Psychiatric, & Specialty Hospitals | 49.5% | 41.5% | 48.2% | 43.0% | 44.3% |
| | Pharmacies & Drug Stores | 58.4% | 42.9% | 51.4% | 43.0% | 49.7% |
| | Veterinary Services | 49.6% | 39.1% | 51.6% | 38.1% | 54.8% |
| | Nursing Care & Child Day Care | 50.4% | 47.3% | 58.8% | 49.2% | 63.8% |
| | Funeral Homes & Services | 45.3% | 41.5% | 45.9% | 38.6% | 39.8% |
| Travel Locations | Car and Other Vehicle Rental | 66.4% | 33.9% | 50.6% | 43.3% | 37.3% |
| | Hotels & Motels | 39.4% | 58.0% | 64.1% | 54.8% | 54.2% |
| Judicial Services | Correctional Institutions | 64.8% | 46.1% | 42.9% | 40.9% | 46.6% |
| | Police Stations | 40.9% | 36.4% | 45.5% | 27.3% | 63.6% |
| Miscellaneous | Libraries & Archives | 55.2% | 38.8% | 50.7% | 39.5% | 44.2% |
| | Religious Organizations | 46.2% | 63.2% | 62.3% | 60.1% | 57.3% |
| | Barber Shops, Beauty & Nail Salons | 50.3% | 46.3% | 44.4% | 38.7% | 43.3% |
| | Movie Theaters | 60.5% | 58.8% | 66.0% | 56.1% | 64.9% |
| | Fitness & Recreational Centers (Gyms) | 55.0% | 48.7% | 50.4% | 52.7% | 51.1% |

*Table 2 Percent of Difference in Differences less than 0 for each period*

**Appendix**

| Main Category | Name | No. locations in high-income zip codes | No. locations in low-income zip codes |
|---|---|---|---|
| Food Services | Full-Service Restaurants | 231 | 192 |
|  | Limited-Service Restaurants | 142 | 87 |
| Essential Goods/Services | Meat, Seafood, & Fruit/Vegetable Markets | 67 | 18 |
|  | Supermarkets & Grocery Stores | 164 | 124 |
|  | Gasoline Stations | 197 | 155 |
|  | Plumbing, Heating, & A.C. Contractors | 28 | 18 |
| Medical Locations | Offices of Physicians | 126 | 93 |
|  | General Medical & Surgical Hospitals | 23 | 43 |
|  | General, Surgical, Psychiatric, & Specialty Hospitals | 34 | 46 |
|  | Pharmacies & Drug Stores | 99 | 65 |
|  | Veterinary Services | 106 | 50 |
|  | Nursing Care & Child Day Care | 149 | 73 |
|  | Funeral Homes & Services | 30 | 35 |
| Travel Locations | Car and Other Vehicle Rental | 30 | 12 |
|  | Hotels & Motels | 111 | 118 |
| Judicial Services | Correctional Institutions | 21 | 44 |
|  | Police Stations | 11 | 4 |
| Miscellaneous | Libraries & Archives | 75 | 65 |
|  | Religious Organizations | 238 | 193 |
|  | Barber Shops, Beauty & Nail Salons | 115 | 72 |
|  | Movie Theaters | 40 | 19 |
|  | Fitness & Recreational Centers (Gyms) | 156 | 69 |

*Table A 1 Relevant NAICS code and number of locations in each type of zip code*

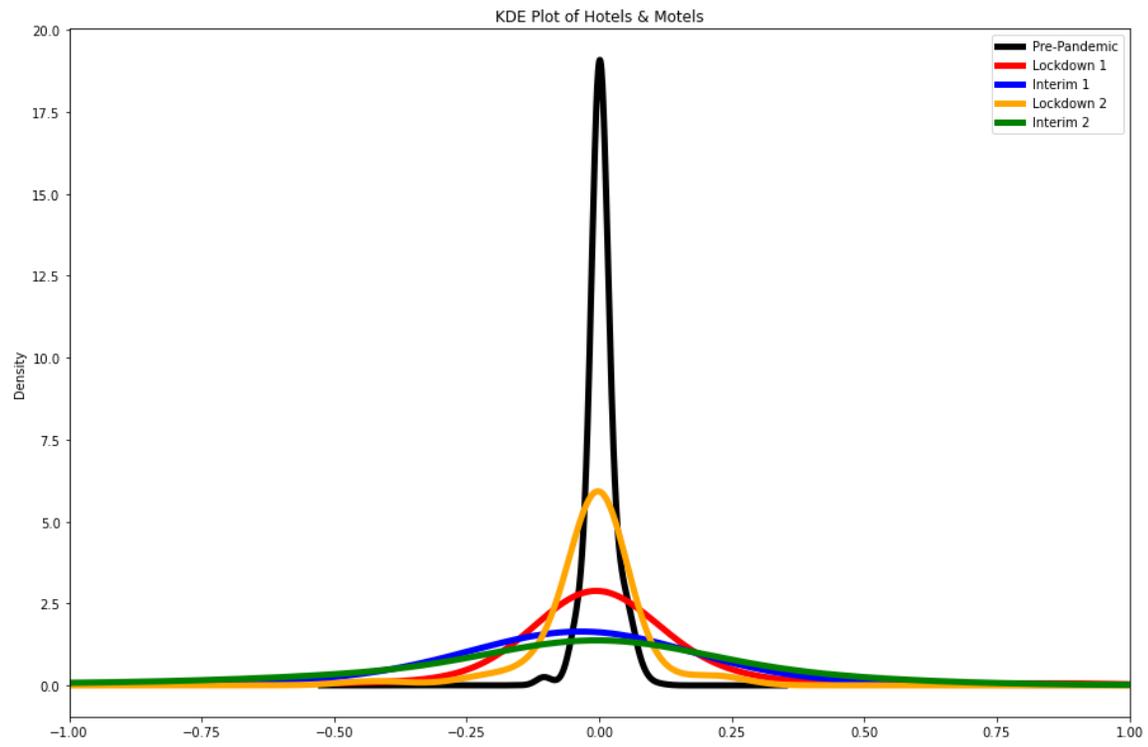

*Figure A 1 Relative changes in the number of visits between low- and high-income zip codes for Hotels & Motels*

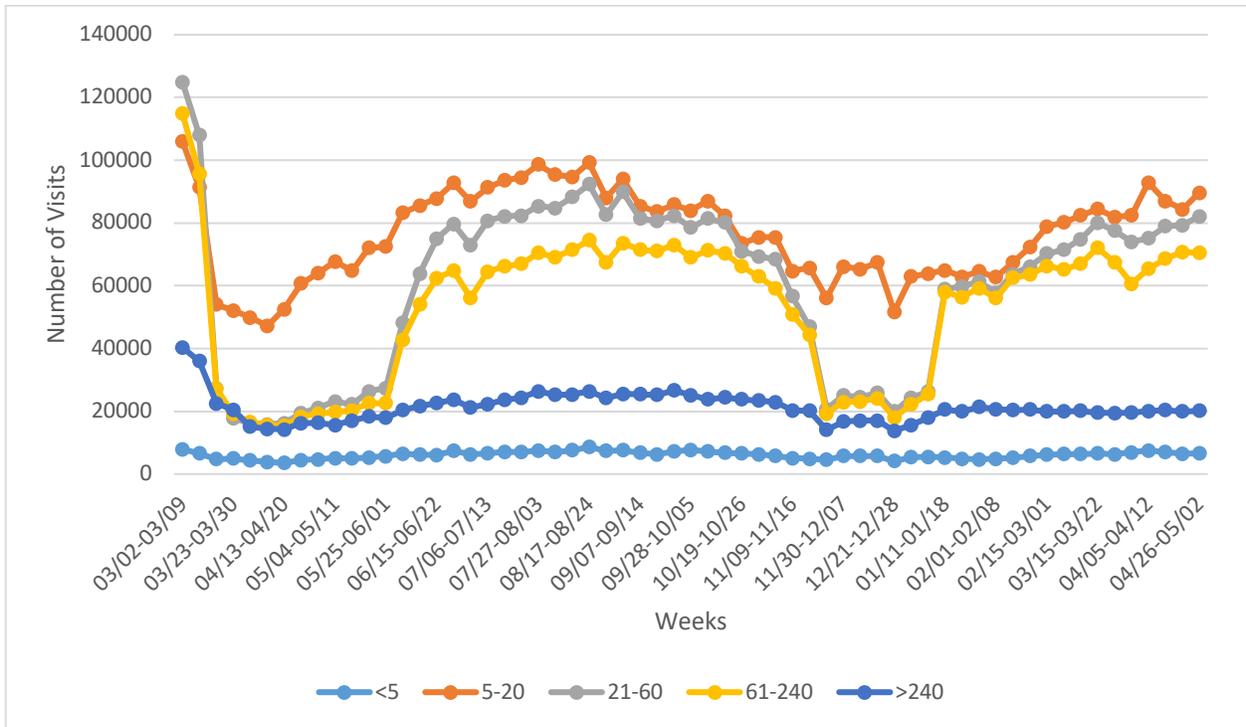

*Figure A 2 Visits to Full-Time Restaurants by Duration (minutes) from March 2020 to May 2021*

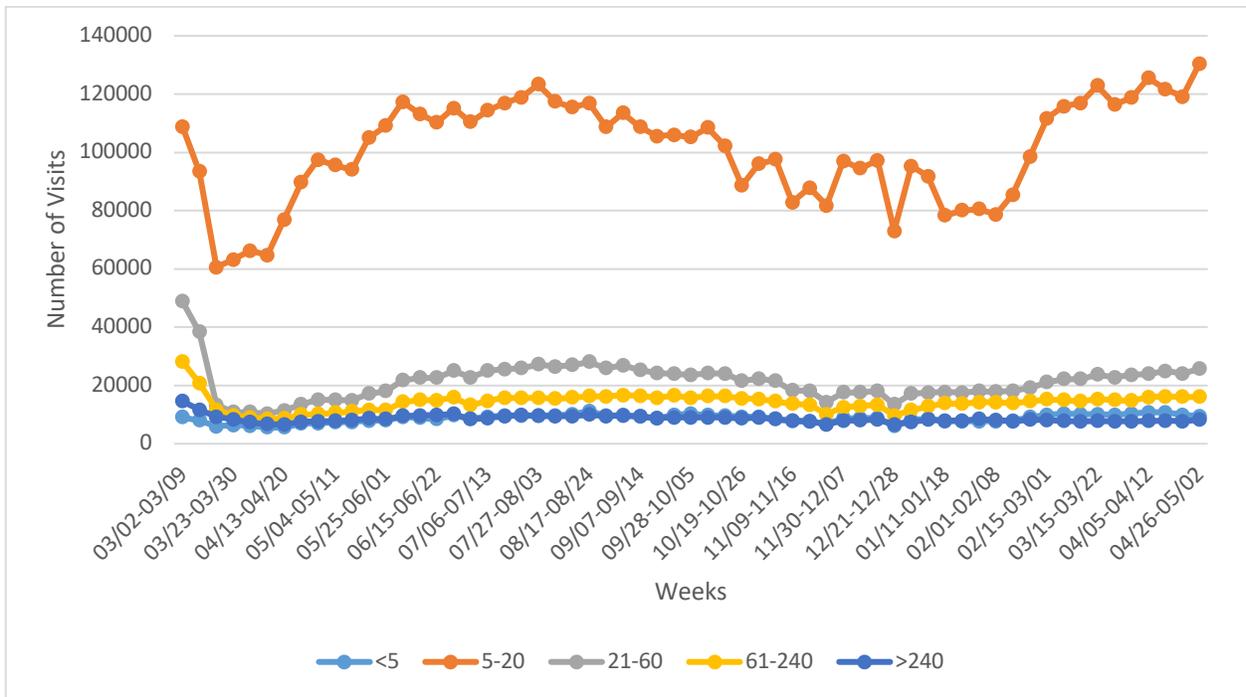

*Figure A 3 Visits to Limited-Service Restaurants by Duration (minutes) from March 2020 to May 2021*

ᵃ There is a lot of consistency in urban mobility at different hours, days, and weeks[12,13].
ᵇ Sometimes, it is not upto the individual: during the pandemic, many low-income workers were classified as doing "critical" work, and as such, were not allowed to work from home.
ᶜ Bundorf et al. (2021)[7] present evidence of large activity reductions in the presence of lockdowns in the U.S.: "40 percent of people reported reducing their activity by a lot for grocery shopping, while 79 percent of people reported reducing their activity by a lot for restaurant visits, consistent with grocery shopping being a more essential activity." Galeazzi et al. (2021)[14] find for France, Italy and UK, lockdowns create "smallworldness —i.e., a substantial reduction of long-range connections in favor of local paths." Cronin and Evans (2020)[15] study foot traffic data for the US and find self-regulating behavior on the part of customers resulting from the changed calculus explains more than three-quarters of the decline in foot traffic in most industries; restrictive regulation explains half the decline.
ᵈ Shin et al. (2021)[16] report significant changes in foot-traffic behavior among Seoul residents. They find reduced visitations are driven by temporary business closures rather than citizens' risk-avoidance behavior.
ᵉ Apedo-Amah et al. (2020)[17] document the severe, widespread, and persistent negative impact on sales across firms worldwide.
ᶠ Fernández-Villaverde and Jones (2020)[18] use Google Mobility data and argue it has several key advantages over sales or GDP-related measures: available at a daily frequency rather than quarterly or monthly; reported with a very short lag; and available at a very disaggregated geographic level. Kim et al. (2022)[19] use individual-level monthly panel data of Singaporeans and find households with above-median net worth (who presumably reside in high-income neighborhoods) reduced their spending more than households with below-median net worth.
ᵍ Target 11.5 of the SDG 11 asks governments to "substantially decrease the direct economic losses relative to global gross domestic product caused by disasters, including water-related disasters, with a focus on protecting the poor and people in vulnerable situations."
ʰ Although device-level demographic data cannot be collected, SafeGraph estimates which census block group the device owner's home belongs to. The data is well-sampled across household income averages, educational attainment, and demographic categories, according to the aggregate summary of SafeGraph data and the characteristics of a Census block group[20].
ⁱ Strictly speaking, SafeGraph's sample is not truly random and is subject to sampling errors. The company has provided evidence that the sample correlates highly with U.S. census data.
ʲ The mode of travel to a POI is not recorded in SafeGraph data, only the number of visits to it. We do not know if a specific person visited a particular POI, A, in 2019 and did or did not visit A in 2020. We also have no data on spending at a POI, only the amount of time spent. Also, our unit of analysis is the zip code.
ᵏ We look at the number of visits by duration and find that visits of less than five minutes, which include pick-up and delivery orders, did not increase significantly during the lockdown. Unfortunately, we are unable to distinguish between delivery orders and pick-up orders, as a single driver can pick up multiple delivery orders. The change is plotted in the appendix, Figures A2-A3.
ˡ We also conducted a sensitivity analysis by changing the thresholds for high- and low-income zip codes from one-third to one-fourth. In that case, only 216 zip codes were reclassified as high-income and 216 as low-income. We also changed the threshold from one-third to two-fifths, so there were 344 high-income and 344 low-income zip codes. After constructing distributions based on the reclassifications, we found the results were similar to the baseline one-third threshold, the one we chose to report.
ᵐ For the first re-opening period, restaurants, bars, theaters, gyms, amusement parks, and personal care services faced an extension of temporary closure (starting from June 2020). But restaurants were able to provide outdoor service provided they met some requirements. Order 20-63 allowed bars and restaurants to serve food outdoors beginning on June 1, 2020. So there were periods when indoor dining was prohibited, but outdoor dining was allowed. However, regardless of the median income of the zip code, everyone in Minnesota was subject to the same order. To our knowledge, no restrictions differed by location (e.g., streets).